\documentclass[conference]{IEEEtran}
\ifCLASSINFOpdf
  \usepackage[pdftex]{graphicx}
\else
\fi
\usepackage{amsmath}
\usepackage{amsfonts}
\usepackage{printlen}  

\begin{document}
\title{Learning Continuous Receive Apodization Weights via Implicit Neural Representation \\ for Ultrafast ICE Ultrasound Imaging}


\author{\IEEEauthorblockN{R\'emi Delaunay\IEEEauthorrefmark{1},
Christoph Hennersperger and
Stefan Wörz}
\IEEEauthorblockA{LUMA Vision, Beech Hill Road, Dublin, Ireland\\ \IEEEauthorrefmark{1} Corresponding author, email: remi.delaunay@lumavision.com}
}

\maketitle

\begin{abstract}
Ultrafast intracardiac echocardiography (ICE) uses unfocused transmissions to capture cardiac motion at frame rates exceeding 1 kHz. While this enables real-time visualization of rapid dynamics, image quality is often degraded by diffraction artifacts, requiring many transmits to achieve satisfying resolution and contrast. To address this limitation, we propose an implicit neural representation (INR) framework to encode complex-valued receive apodization weights in a continuous manner, enabling high-quality ICE reconstructions from only three diverging wave (DW) transmits. Our method employs a multi-layer perceptron that maps pixel coordinates and transmit steering angles to complex-valued apodization weights for each receive channel. Experiments on a large \textit{in vivo} porcine ICE imaging dataset show that the learned apodization suppresses clutter and enhances contrast, yielding reconstructions closely matching 26-angle compounded DW ground truths. Our study suggests that INRs could offer a powerful framework for ultrasound image enhancement.
\end{abstract}
 
\section{Introduction}

% context and limitations
Ultrafast ultrasound imaging enables high frame rates by insonifying the entire imaging field with a single unfocused wavefront, i.e., a plane or diverging wave, rather than using sequential focused beams \cite{tanter2014ultrafast}. This strategy allows frame rates to exceed 1,000 frames per second, making it highly suitable for applications requiring rapid temporal resolution, such as cardiac and vascular imaging \cite{villemain2020ultrafast}. However, the absence of transmit focusing compromises spatial resolution and leads to the formation of side lobe artifacts. Coherent compounding addresses this limitation by summing multiple unfocused transmissions acquired at different steering angles, improving image quality at the expense of temporal resolution \cite{montaldo2009coherent}.

Delay-and-sum (DAS) beamforming is the most widely used approach for ultrasound image reconstruction, where echoes received by each transducer element are time-aligned and summed to form each pixel of the image. While its straightforward implementation and low computational cost make it attractive, the resulting image quality is often suboptimal~\cite{perrot2021so}. To mitigate clutter and side lobes, DAS is frequently combined with apodization, applied after time-of-flight correction, where the received signals are weighted (typically using predefined window functions such as Hann or Tukey) to modulate their amplitude across the transducer elements~\cite{hoen1982aperture}.

To further improve image quality, advanced beamforming techniques have been developed, notably adaptive beamforming methods that weight the receive, time-of-flight corrected channel signals based on the spatial and statistical properties of the echo data, rather than relying on fixed apodization windows. Examples include coherence-based approaches~\cite{camacho2009phase}, minimum variance (MV) beamforming~\cite{synnevag2007adaptive}, and iterative maximum a posteriori (iMAP) methods~\cite{chernyakova2019imap}. These techniques have demonstrated substantial gains over DAS beamforming, enhancing contrast and resolution, but often at the cost of increased computational complexity. More recently, deep learning-based methods have emerged, either trained under supervision from high-quality reference images \cite{perdios2021cnn} or inspired by model-based strategies \cite{luijten2020adaptive,van2019deep}. These methods aim to combine the flexibility of adaptive beamformers with the efficiency and generalization ability of data-driven models, opening new possibilities for real-time, high-quality ultrasound imaging.

Implicit Neural Representations (INRs) are a class of neural network architectures that encode continuous functions directly within their weights. Initially popularized in computer graphics through applications such as Neural Radiance Fields (NeRF) \cite{mildenhall2021nerf}, INRs have since demonstrated strong performance in representing other complex signals such as images and audio signals \cite{sitzmann2020siren}. Recently, INRs have been applied to ultrasound imaging in various contexts, including ultrasound image synthesis (Ultra-NeRF)~\cite{wysocki2024ultra}, compact representations of plane wave acquisition sequences~\cite{monvoisin2024compact}, and speed-of-sound estimation~\cite{byra2024implicit}. Unlike conventional neural networks, which operate on discrete grid-based data, INRs represent signals continuously, enabling flexible and resolution-independent modeling of the target quantity.

In this paper, we propose a novel learning-based adaptive beamforming method using an INR-based framework. Our approach trains an MLP to learn a continuous representation of complex-valued apodization weights conditioned on pixel coordinates and transmit steering angles. By embedding apodization into an INR, the network can predict spatially adaptive, angle-aware weights for each receive channel from the time-delayed I/Q data. The model is trained to reconstruct high-quality intracardiac echocardiography (ICE) images from only three diverging wave transmissions, using high-quality images compounded with 26 transmissions as ground truth. Our method enables data-efficient, high-quality reconstruction while substantially reducing the number of required transmit events.

\begin{figure*}[!t]
\centering
\includegraphics[width=\textwidth]{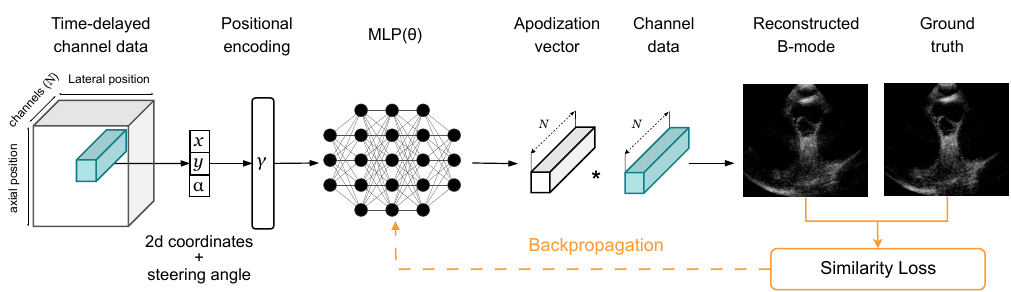}
\caption{Method overview. The 2D pixel coordinates and steering angle extracted from the ultrasound image are transformed using complex sinusoidal positional encoding and fed into an MLP-based INR network. The INR predicts complex-valued apodization weights, which are applied to the corresponding time-delayed receive channel data during beamforming and image reconstruction. }
\label{fig_overview}
\end{figure*}

\section{Methods}
\subsection{Complex-valued implicit network}
The proposed method is an implicit neural representation that enhances ultrasound image quality by learning a continuous mapping from spatial coordinates and steering angles to a set of receive apodization values. An overview of the method is presented in Fig. \ref{fig_overview}. Our implicit representation consists of a multi-layer perceptron (MLP) and can be defined by the following mapping function:
 
\begin{equation}
f_\theta(\mathit{x, z, \alpha}) = \mathbf{w},
\end{equation}

\noindent where $(\mathit{x, z, \alpha}) \in \mathbb{R}^3$ is a 3D coordinate vector representing the lateral and axial pixel positions $(x, z)$ and the steering angle $\alpha$ of the transmit event. The network outputs a complex-valued apodization vector $\mathbf{w} \in \mathbb{C}^{N}$, where $N$ is the number of receive channels. To enable the MLP to capture high-frequency details in the complex-valued space, the input coordinate vector is first transformed using a complex sinusoidal positional encoding function $\gamma(\cdot)$, which maps each scalar real component to a higher-dimensional, complex-valued space:

\begin{equation}
\label{eq_pe}
\gamma(p) = \left[ e^{i 2^0 \pi p},\; e^{i 2^1 \pi p},\; \ldots,\; e^{i 2^{L-1} \pi p} \right],
\end{equation}

\noindent where $p$ corresponds to the input coordinate, and {L} is the embedding length. This complex-valued positional encoding extends the original NeRF formulation \cite{mildenhall2021nerf} by replacing the real-valued sine and cosine basis functions with complex exponentials. This implicitly captures both sine and cosine components through Euler’s identity, while naturally transforming real-valued coordinates into complex-valued representations compatible with the complex MLP. By operating directly in the complex domain, the encoding enables the network to more effectively model high-frequency spatial variations and angular dependencies that are inherent in ultrasound signal formation.

\subsection{Image Reconstruction}

We performed DAS beamforming on In-phase/Quadrature (IQ) demodulated data to reconstruct the ultrasound images. For each transmit event, the IQ signal \(\mathbf{s}_n(t)\) from each receive channel \(n\) is time-of-flight corrected by interpolating the signal at the round-trip propagation delay \(\tau_{n,p}\) corresponding to each image pixel \(p = [x, z]\):

\begin{equation}
s_n[x, z] = \mathbf{s}_n(\tau_{n,p}),
\end{equation}

\noindent where \(\tau_{n,p}\) is computed based on the known geometry and speed of sound, and \(s_n[x, z]\) represents the time-delayed complex I/Q signal for pixel \([x, z]\). The predicted complex apodization weights \(\mathbf{w}_p \in \mathbb{C}^N\), obtained from the INR, are then applied to form the beamformed image as:

\begin{equation}
y_{\text{DAS}}[x, z] = \sum_{n=1}^{N} w_n[x, z] \cdot s_n[x, z],
\end{equation}

\noindent where \(w_n[x, z]\) is the learned complex apodization weight for the \(n\)-th channel at pixel \([x, z]\). This operation is repeated for all transmit events, and the resulting images are summed to form the final compounded frame.

\subsection{Loss function}

The network is trained in a supervised manner by comparing the B-mode reconstructed image \(I_{\mathrm{pred}}\), compounded with three transmit events, to a high-quality ground truth image \(I_{\mathrm{gt}}\) generated by compounding 26 transmits. The training loss $\mathcal{L}$ combines both mean squared error (MSE) and structural similarity index measure (SSIM) metrics, formulated as

\begin{equation}
\mathcal{L} = \beta \cdot \mathrm{MSE}(I_{\mathrm{pred}}, I_{\mathrm{gt}}) + (1 - \beta) \cdot \bigl(1 - \mathrm{SSIM}(I_{\mathrm{pred}}, I_{\mathrm{gt}})\bigr),
\end{equation}

\noindent where \(\beta = 0.5\) is chosen empirically to balance pixel-wise accuracy and structural fidelity. Both predicted and ground truth signals undergo B-mode conversion—which includes envelope detection followed by log-compression—ensuring that the network optimizes perceptual image quality by directly comparing their final visual representations rather than the raw complex-valued data.

\section{Experimental Setup}

\subsection{\textit{In Vivo} ICE Dataset}

The \textit{in vivo} dataset used in this study was acquired from porcine subjects using a custom-built intracardiac echocardiography (ICE) catheter comprising 64 elements operating at a central frequency of 6~MHz, developed by LUMA Vision\footnote{LUMA Vision Ltd., Dublin, Ireland, www.lumavision.com}. During acquisition, the ICE catheter was navigated within the heart to capture a wide variety of anatomical views. A total of 9{,}780 two-dimensional diverging wave (DW) acquisitions were collected with varying parameters, including virtual source distances (10--20~mm) and sector angles (45--90$^\circ$), with an imaging depth of 90~mm.

Ground truth images were reconstructed using DAS beamforming with rectangular receive apodization (\( \textbf{w} = \mathbf{1}^N \)), combining 26 steered DW transmits equally spaced across the sector angle. For training, the model was provided with individual DW acquisitions, using corresponding pixel coordinates and steering angles as inputs. During testing, the network output was evaluated on a three-transmit configuration, using the leftmost, center, and rightmost steering angles to reflect practical constraints in high frame-rate imaging. The INR predicted a set of complex apodization weights for each of the three transmits, which were then used to coherently compound the final image. Baseline low-quality inputs were constructed from the same three transmits using rectangular apodization. The dataset was partitioned into training, validation, and test sets with a 60:20:20 ratio.

\subsection{Implementation and Training}

The INR model architecture follows the NeRF-style coordinate-based design \cite{mildenhall2021nerf} and is implemented as a MLP with six complex-valued fully connected layers, each containing 128 hidden units. A skip connection is introduced at the fourth layer, feeding the embedded input forward. All layers are followed by a modulus ReLU (modReLU) activation function \cite{barrachina2023theory}, except for the final layer, which uses a complex-valued sigmoid activation \cite{smith2023complex}. The complex sigmoid ensures that the predicted apodization weights are constrained between 0 and 1 in both the real and imaginary components, effectively regularizing the output and promoting stable convergence during training.

Spatial coordinates and steering angles were normalized to the \([0, 1]\) range and passed through complex sinusoidal positional encodings with an embedding size of 10 (see Eq. \ref{eq_pe}). The model was trained using the Adam optimizer over 100 epochs, with a batch size of 4 and an initial learning rate of 1e-4. All experiments were conducted using PyTorch on an NVIDIA RTX A6000 GPU.

To evaluate image quality, we compared the predicted and baseline reconstructions against the ground truth high-quality images using structural similarity (SSIM) and peak signal-to-noise ratio (PSNR) metrics.

\section{Results}
\begin{figure}[!t]
\centering
\includegraphics[width=0.48\textwidth]{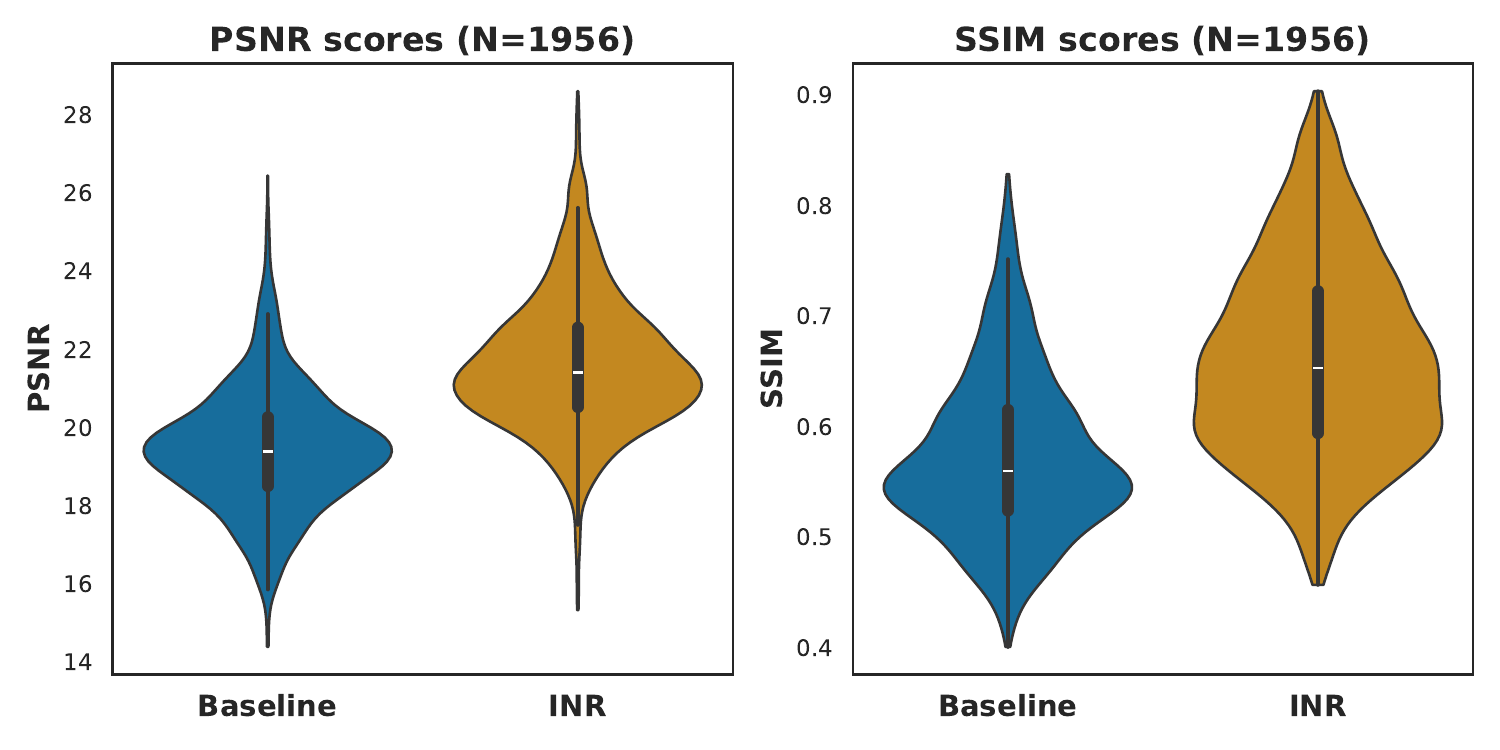}
\caption{Violin plots showing the distribution of PSNR and SSIM scores on the testing \textit{in vivo} ICE dataset. Black rectangles indicate boxplots, and white lines represent the median values.}
\label{fig_boxplot}
\end{figure}

A violin figure showing the distribution of SSIM and PSNR scores for the baseline and method's output is showed in Fig. \ref{fig_boxplot}. Compared to DAS with three transmits, our method improved average PSNR scores from 19.44 ($\pm 1.56$) dB to 21.62 ($\pm 1.79$) \\dB and average SSIM scores from 0.57 ($\pm 0.07$) to 0.66 ($\pm 0.09$) for the inference of an \textit{in vivo} ICE dataset comprising 1956 2D images.

\begin{figure}[!t]
\centering
\includegraphics[width=0.48\textwidth]{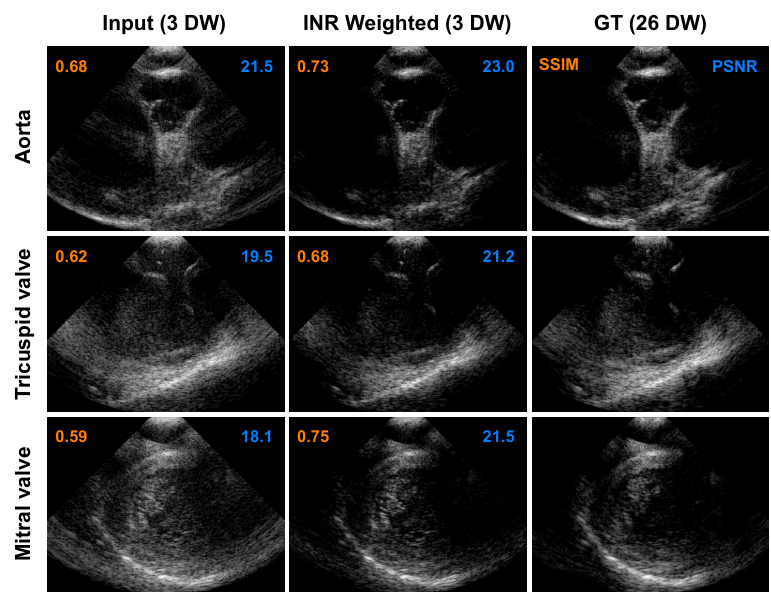}
\caption{Qualitative comparison on the \textit{in vivo} ICE dataset between the baseline (3 transmits), the baseline with predicted INR apodization weights, and the ground truth (26 transmits). Each row shows different anatomical views: aorta, tricuspid valve, and mitral valve.}
\label{fig_result}
\end{figure}

Figure~\ref{fig_result} shows qualitative comparisons of different anatomic views taken from a porcine subject (mitral valve, tricuspid valve and aorta). The INR reconstruction effectively suppresses clutter and side lobes artifacts visible in 3-transmit DAS, yielding images that closely match the 26-angles compounded ground truth references. 

\section{Discussion}

In this work, we demonstrated the feasibility of encoding pixel-wise apodization weights into an INR. In addition, we introduced a new version of sinusoidal positional encoding that converts real coordinates and steering angles into a higher-dimensional, complex-valued representation. This enables our network to be fully complex-valued, making it ideal for modeling IQ signals, which is ubiquitous in the ultrasound imaging field. Our results show that the learned apodization weights can map low-quality inputs—reconstructed from only three DW events—into high-quality images, effectively suppressing side lobes and clutter noise. These findings suggest that INRs offer a powerful framework for ultrasound image enhancement and could be extended to learn more advanced mappings, such as adaptive or regularization-based beamforming strategies.

Beyond image quality improvements, the continuous nature of the INR enables natural generalization to arbitrary spatial grids and acquisition settings, providing a flexible alternative to other network architectures that rely on discretized grids. Furthermore, once trained, inference is needed only once for a given set of pixel coordinates and steering angles, so processing does not incur additional overhead, unlike traditional adaptive approaches that are iterative and computationally intensive.

A current limitation of our framework is that the learned representation is not dynamically conditioned on the input channel data, which can reduce robustness to out-of-distribution inputs. This drawback is inherent to INR methods, though recent work on conditional learning provides promising directions \cite{mehta2021modulated}. Future extensions could integrate such strategies, enabling the model to dynamically adapt apodization weights to input-specific features.

\newpage
\bibliographystyle{IEEEtran}
\bibliography{IEEEabrv,references.bib}

\end{document}